\documentclass[aps,prb,twocolumn,superscriptaddress]{revtex4-2}
\usepackage{amsmath}
\usepackage{amsfonts}
\usepackage{amssymb}
\usepackage{graphicx}
\usepackage{xcolor}
\usepackage[colorlinks=True,citecolor=myblue,linkcolor=myorange,urlcolor=mygreen]{hyperref}
\usepackage{hypcap}
\usepackage{braket}
\usepackage{bm}
\usepackage{bbm}
\usepackage[export]{adjustbox}
\usepackage{verbatim}

\definecolor{myblue}{RGB}{0,0,130}
\definecolor{myorange}{RGB}{130,50,0}
\definecolor{mygreen}{RGB}{0,130,0}

\begin{document}

\title{Resonant energy scales and local observables in the many-body localised phase}

\author{Samuel J. Garratt}
\email{sjgarratt@berkeley.edu}
\affiliation{Department of Physics, University of California, Berkeley, California 94720, USA}
\author{Sthitadhi Roy}
\email{sthitadhi.roy@icts.res.in}
\affiliation{International Centre for Theoretical Sciences, Tata Institute of Fundamental Research, Bengaluru 560089, India}
\affiliation{Rudolf Peierls Centre for Theoretical Physics, Oxford University, Parks Road, Oxford OX1 3PU, UK}
\affiliation{Physical and Theoretical Chemistry, Oxford University, South Parks Road, Oxford OX1 3QZ, UK}

\begin{abstract}
We formulate a theory for resonances in the many-body localised (MBL) phase of disordered quantum spin chains in terms of local observables. A key result is to show that there are universal correlations between the matrix elements of local observables and the many-body level spectrum. This reveals how the matrix elements encode the energy scales associated with resonance, thereby allowing us to show that these energies are power-law distributed. Using these results we calculate analytically the distributions of local polarisations and of eigenstate fidelity susceptibilities. The first of these quantities characterises the proximity of MBL systems to noninteracting ones, while the second highlights their extreme sensitivity to local perturbations. Our theoretical approach is to consider the effect of varying a local field, which induces a parametric dynamics of spectral properties. We corroborate our results numerically using exact diagonalisation in finite systems.
\end{abstract}

\maketitle

\section{Introduction}
If a many-body quantum system is isolated from its environment, it does not necessarily thermalise. In systems that do thermalise, observables are at late times insensitive to the initial conditions \cite{deutsch1991quantum,srednicki1994chaos,rigol2008thermalization,dallesio2016from,deutsch2018eigenstate}, which is to say that their fluctuations are small. This behaviour is reflected in the resemblance between many-body eigenstates and high-dimensional random vectors, which appear featureless to any local probe. The alternative to this ergodic behaviour, arising in the presence of strong disorder, is many-body localisation~\cite{gornyi2005interacting,basko2006metal,oganesyan2007localization,nandkishore2015manybody,abanin2019colloquium}. In the many-body localised (MBL) phase there is memory of local observables even at infinite times. In MBL quantum spin chains, for example, eigenstates feature large local polarisations, and these polarisations vary dramatically both in space and over the ensemble of disorder realisations. 

Large fluctuations are also found in dynamics. For this reason, averages often fail to represent accurately the behaviour of individual MBL systems. A striking example is found in temporal correlations of local observables, and the associated spectral functions \cite{serbyn2017thouless}. To characterise the MBL phase, it is therefore necessary to develop a theory for the statistical properties of these quantities. In quantum-mechanical systems it is natural to do so in the spectral representation, and there the question is whether and how the matrix elements of local observables are related to the level spectrum. 

In this work we show that there are universal correlations between these quantities, and that these correlations are characterised by a power-law distribution of energy scales. From these correlations we determine the statistics of local polarisations, and find that these too are broadly distributed. The fluctuations of physical properties are reflected in the extreme sensitivity of eigenstates to local perturbations, and to investigate this we analytically determine the distribution of eigenstate fidelity susceptibilities in the MBL phase. This quantity has recently been investigated extensively as a probe of integrability breaking and of the onset of quantum chaos \cite{sierant2019fidelity,maksymov2019energy,sels2020dynamical,crowley2021constructive,pandey2020adiabatic,leblond2020universality,crowley2021partial}.

The calculations described above are unified by the concept of a many-body resonance~\cite{gopalakrishnan2015low,villalonga2020eigenstates,crowley2021constructive,morningstar2021avalanches,garratt2021local}. Many-body resonances are generalisations of the Mott resonances that arise in Anderson insulators \cite{mott1968conduction,berezinskii1974kinetics,ivanov2012hybridization}, and their presence demarcates MBL systems from their noninteracting counterparts. These resonances control the low-frequency dynamics in the MBL phase \cite{gopalakrishnan2015low,colmenarez2019statistics,crowley2021constructive,garratt2021local}, and their local structure is central in our understanding of both its stability~\cite{imbrie2016diagonalization,crowley2021constructive} and its breakdown~\cite{deroeck2017many,deroeck2017stability,morningstar2021avalanches}. One possibility is to describe resonances in terms of local integrals of motion (LIOM)~\cite{serbyn2013local,huse2014phenomenology}, a hypothesised set of quasi-local operators that commute with one another and with the evolution operator. However, this approach creates a barrier to quantitative investigations because LIOM cannot be defined uniquely \cite{imbrie2017local,ros2015integrals,chandran2015constructing,rademaker2016explicit,pekker2017fixed}. Additionally, schemes intended to construct them inevitably break down in the vicinity of the transition to the ergodic phase. It is therefore essential to develop a more complete understanding of resonances that does not rely on LIOM, and is instead based on quantities that are both computationally accessible and unambiguous. With this motivation, in this work we instead formulate our theory in terms of the matrix elements of local observables. These matrix elements can be calculated using standard numerical techniques.

This paper is organised as follows. In Sec.~\ref{sec:resonances} we describe the idea of a many-body resonance. Following this in Sec.~\ref{sec:parametric} we discuss how their local structure can be captured analytically. This leads us to the central result of this work in Sec.~\ref{sec:energy}. There we show how the matrix elements of observables are related to energy splittings, and calculate the distribution of energy scales governing dynamics in the MBL phase. The form of this distribution is elucidated through a resonance counting argument in Sec.~\ref{sec:counting}. Using these results, in Sec.~\ref{sec:polarisations} we calculate the distribution of polarisations, and in Sec.~\ref{sec:perturbations} that of fidelity susceptibilities. We summarise our results in Sec.~\ref{sec:summary}. 

\begin{figure}
\includegraphics[width=\linewidth]{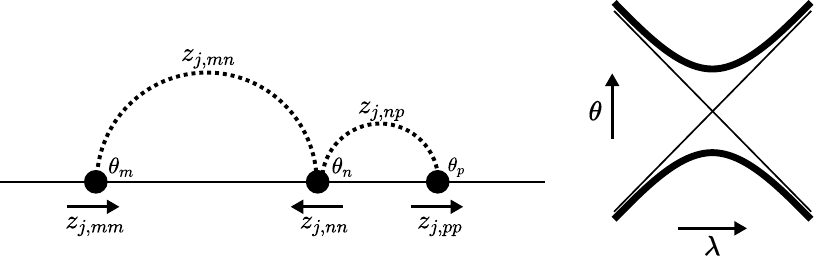}
\caption{Parametric dynamics induced by varying a local field $h_j(\lambda)=h_j+\lambda$, which couples to the local operator $\sigma_j$ at site $j$. The parameter $\lambda$ can be viewed as a fictitious time, and in the case of Floquet dynamics we denote by $e^{i\theta_n}$ the eigenvalues of the Floquet operator. Left: Three levels at a particular $\lambda$, with the $\theta$ axis horizontal. The diagonal matrix elements $z_{j,nn}$ of $\sigma_j$ determine level velocities $\partial_{\lambda}\theta_n$, and off-diagonal matrix elements $z_{j,nm}$ increase in magnitude as levels approach resonance. Right: Thin lines show the level dynamics for a system with decoupled degrees of freedom, where an exact level crossing occurs as $\lambda$ is varied, and thick lines correspond to the interacting case, where the crossing is avoided and corresponds to a resonance.}
\label{fig:parametric}
\end{figure}

\section{Local resonances}\label{sec:resonances}

Before discussing resonances in MBL spin chains it is helpful to first outline the origins and structure of Mott resonances in non-interacting Anderson insulators. We can imagine starting from a set of perfectly localised single-particle eigenstates on a lattice whose sites have random energies. Introducing weak hopping between neighbouring sites, typical eigenstates acquire exponentially-decaying tails in real space, but remain well-localised. However, some pairs of sites will be sufficiently close in energy that they are strongly hybridised by the hopping, and the resulting eigenstates will have significant weight on each of the two sites involved. These resonances can be described approximately as two level systems \cite{mott1968conduction,ivanov2012hybridization}.

The MBL systems of interest here are spin chains in random local fields, and with weak short-range interactions on energy scale $J$. For $J=0$, these are systems of decoupled spins, and we denote by $(h_j/2) \sigma_j$ the random local field operators, where $j=1 \ldots L$ labels the sites. The field strengths $h_j > 0$ are of order unity and the local operators $\sigma_j$ have eigenvalues $\pm 1$. For $J=0$, the eigenstates $\ket{n}$ of the evolution operator satisfy $\sigma_j \ket{n}=s_{n,j}\ket{n}$ with $s_{n,j}=\pm 1$, and here the LIOM are simply the operators $\sigma_j$. For small $J \neq 0$ it is expected that the LIOM are dressed by the interaction, and thereby develop exponentially-decaying tails in real space~\cite{serbyn2013local,huse2014phenomenology}. Their eigenvalues could then be used to label the eigenstates. In this sense the eigenstates locally resemble those in the decoupled system. On the other hand, we know that if a pair of eigenstates are for $J=0$ sufficiently close in energy, then when switching on the interaction $J \neq 0$ these states will be strongly hybridised. We refer to this situation as a many-body resonance.

The stability of the MBL phase relies on the fact that resonances are not simple two level systems. Where there are two-dimensional resonant subspaces, these are embedded locally within pairs of many-body eigenstates. Previously this feature has been captured by appealing directly to pictures based on LIOM \cite{gopalakrishnan2015low,crowley2021constructive}, and by considering the spectral properties of evolution operators defined on finite subregions \cite{imbrie2016diagonalization,garratt2021local}, although each of those approaches suffers from a degree of ambiguity (this problem can of course be avoided by studying resonances on the scale of the system size \cite{villalonga2020eigenstates,morningstar2021avalanches}). Accounting for locality is essential because, even deep within the MBL phase, eigenstates are involved in an extensive number of local resonances. Here we describe local resonances in terms of the matrix elements of local observables, thereby isolating the participating degrees of freedom. This will allow us to describe the two-dimensional resonant subspaces in terms of local properties of many-body eigenstates.

\section{Parametric dynamics}\label{sec:parametric}

Our theoretical considerations are based on the effect of parametric variations of the disorder, which can be viewed as inducing dynamics of the spectral properties in a fictitious time; see Fig.~\ref{fig:parametric} for a schematic visual. The idea is particularly powerful in this setting because the avoided crossings that arise under parametric dynamics can be identified with many-body resonances \cite{garratt2021local}. The parametric approach has a long history in studies of random matrices~ \cite{dyson1962brownian}, semiclassical chaos~ \cite{pechukas1983distribution,yukawa1985new,nakamura1986complete}, and disordered conductors~ \cite{szafer1993universal,simons1993universal,simons1993universalities,chalker1996random}, and has more recently been applied in the context of MBL \cite{serbyn2016spectral,filippone2016drude,monthus2016level,monthus2017manybody,maksymov2019energy,de2021intermediate}. A common approximation is to decouple the parametric dynamics of eigenvalues from that of eigenstates, which is reasonable in phases where eigenstates are almost featureless. As in Ref.~\cite{garratt2021local} we do not make this approximation. To understand the structure of resonances it is essential to understand how eigenvalues and eigenstates are coupled with one another. 

Using this idea we will show how a two-dimensional resonant subspace can be described in terms of local properties of many-body eigenstates $\ket{n}$. First observe that for $J=0$ the eigenstates are tensor products of eigenstates of each of the $\sigma_j$, so are perfectly polarised along the field directions: $z_{j,nn} \equiv \braket{n|\sigma_j|n} = s_{n,j}$. Additionally, the off-diagonal matrix elements $z_{j,nm} \equiv \braket{n|\sigma_j|m} = 0$. Since the eigenvalues of the time evolution operator are for $J=0$ determined by signed sums of local fields $\sum_j s_{n,j} h_j$ there is no level repulsion. If we consider a smooth variation of a single local field $h_j$ then for $J=0$ there are many exact level crossings. For small $J \neq 0$, all of these exact level crossings are replaced by avoided crossings, as illustrated in Fig.~\ref{fig:parametric}. Moreover, in the vicinity of an avoided crossing between $\ket{n}$ and $\ket{m}$ induced by varying $h_j$, the off-diagonal matrix element $z_{j,nm}$ is of order unity \cite{garratt2021local}. Since $\sum_m |z_{j,nm}|^2=\sum_n |z_{j,nm}|^2=1$, this implies that $z_{j,nn}$ and $z_{j,mm}$ are suppressed to well below unity, and therefore that near $j$ the states $\ket{n}$ and $\ket{m}$ do not resemble the $J=0$ eigenstates.

There are two classes of variations that must be distinguished. For $J=0$ and focusing a particular level pair, varying the field $h_j(\lambda)=h_j+\lambda$ will lead either to an exact crossing of levels in the case $s_{n,j}=-s_{m,j}$, or otherwise cause them to move parallel to one another in the case $s_{n,j}=s_{m,j}$. While for small $J \neq 0$ the exact crossings are replaced by avoided ones, we do not expect a dramatic difference in behaviour in the other case. Focus now on $\ket{n}$, $\ket{m}$ and a variation $h_j$ that leads to an avoided crossing. We denote by $\omega^*_{j,nm}$ the minimum level separation, which occurs at fictitious time $\lambda^*_{j,nm}$. From perturbation theory in $h_j$ we find
\begin{align}
	\begin{split}
 	\partial_{\lambda}\omega_{nm} &=  y_{j,nm}  \equiv ( z_{j,nn} - z_{j,mm} )/2 \\
	\partial_{\lambda} \ket{n} &= \frac{1}{2} \sum_{p \neq n} \omega_{np}^{-1} z_{j,pn} \ket{p}.
	\end{split}
\label{eq:firstorderPT}
\end{align}
For Floquet systems the factor $\omega_{np}^{-1}$ should be replaced by $i [1-e^{-i\omega_{np}}]^{-1}$, but these coincide for $|\omega_{np}| \ll \pi$. We find for the matrix elements of $\sigma_j$,
\begin{align}
	\partial_{\lambda} z_{j,nm} &= - \omega_{nm}^{-1} z_{j,nm} y_{j,nm}  + \ldots \label{eq:dz}\\
	\partial_{\lambda} y_{j,nm} &= \omega^{-1}_{nm} |z_{j,nm}|^2  + \ldots, \notag
\end{align}
where the ellipses represent contributions depending on $z_{j,np}$ and $z_{j,mp}$ for $p \neq m,n$. These contributions are significant only for states $\ket{p}$ that resonate with $\ket{n}$ or $\ket{m}$ at the site $j$, and in order to make analytic progress we neglect them in the first instance. This approximation that resonances are locally rare will be shown to be self-consistent below. 

Within the approximation of decoupled local resonances we find a closed system of differential equations that describe variations of $\omega_{nm}$ and of the matrix elements in Eq.~\ref{eq:dz}. This set of equations simply describes an avoided crossing in a two-dimensional resonant subspace, but crucially is expressed only in terms of properties of the eigenstates $\ket{n}$ and $\ket{m}$ that are local to the site $j$. The solutions, which we discuss in Appendix~\ref{sec:solution}, are parametrised by three constants of integration. Two of these are $\lambda^*_{j,nm}$ and $\omega^*_{j,nm}$, and the third $R^2_{j,nm} \equiv y_{j,nm}^2 + |z_{j,nm}|^2$ sets the scale of the variation of the matrix elements as the level pair passes through resonance. Since for $J=0$ we here have $R^2_{j,nm}=1$, it is natural to expect that for $J \neq 0$ the constant $R^2_{j,nm}$ is reduced only by an amount of order $J$. It is straightforward to show that $\partial_{\lambda}(\omega_{nm}z_{j,nm})=0$ and
\begin{align}
	|z_{j,nm} \omega_{nm}| = |R_{j,nm} \omega^*_{j,nm}|. \label{eq:Romegastar}
\end{align}
As the splitting $\omega_{nm}$ passes through a minimum, the magnitude of the matrix element $z_{j,nm}$ passes through its maximum value of $R_{j,nm}$, and these two effects cancel with one another. For small $J$ we expect the above to provide an approximate description of avoided crossings between essentially all level pairs, regardless of their separation in the spectrum. For example, the highest energy resonances correspond to $|z_{j,nm}| \sim 1$ for $\omega_{nm}$ that is exponentially larger in $L$ than the mean level separation. 

The conservation of $z_{j,nm}\omega_{nm}$ under variations of $h_j$ within the approximation of decoupled resonances hints at interesting features in the spectral properties. Before discussing these, it is important to bear in mind that even within this approximation each of $z_{j,nm} \omega_{nm}$ and $\omega_{nm}$ depend on $h_{k \neq j}$. We will additionally assume that any variations of $z_{j,nm} \omega_{nm}$ and $\omega_{nm}$ that are induced by those of $h_{k \neq j}$ are uncorrelated. Crucially, we are not assuming that variations of $z_{j,nm}$ and $\omega_{nm}$ are uncorrelated; a change in the field $h_k$ that brings $\ket{n}$ and $\ket{m}$ closer to a resonance that involves the site $j$ will cause $\omega_{nm}$ to decrease and $|z_{j,nm}|$ to increase. Of course, outside of the approximation of decoupled resonances, there are also small variations of $z_{j,nm}\omega_{nm}$ with $h_j$, and we discuss these in Appendix~\ref{sec:threebody}. In summary, within the above approximations $z_{j,nm} \omega_{nm}$ and $\omega_{nm}$ are only weakly correlated with one another.
 
\section{Energy scales}\label{sec:energy}

An important feature of the MBL phase is that the energy scales associated with resonances have an extremely broad distribution \cite{morningstar2021avalanches,garratt2021local}. Our arguments in the previous section show how these energy scales are encoded in the matrix elements $z_{j,nm}$ and in the level separations $\omega_{nm}$. If we write
\begin{align}
	|z_{j,nm}| = \frac{\Omega_{j,nm}}{|\omega_{nm}|}\,,
\label{eq:Omega}
\end{align}
then the ensemble-averaged distribution $p_{\Omega}(\Omega_{j,nm})$ of `resonant energy scales' $\Omega_{j,nm}$ is independent of $\omega_{nm}$ for $\Omega_{j,nm}<|\omega_{nm}|$. Given this distribution, one would have access to the correlations between $z_{j,nm}$ and $\omega_{nm}$, and therefore the real-time dynamics of the local observable $\sigma_j$. For example, the infinite temperature autocorrelation function $C(t) = 2^{-L} \text{Tr}[\sigma_j(t)\sigma_j] = 2^{-L}\sum_{nm}|z_{j,nm}|^2e^{-i\omega_{nm}t}$. The increase of $z_{j,nm}$ with decreasing $\omega_{nm}$ in Eq.~\eqref{eq:Omega} indicates that low frequency oscillations have large amplitudes. On the other hand, the slow decrease of $z_{j,nm}$ with increasing $\omega_{nm}$ indicates that the potential for resonance is felt even for $|\omega_{nm}| \gg \Omega_{j,nm}$.

\begin{figure}
	\includegraphics[width=\linewidth]{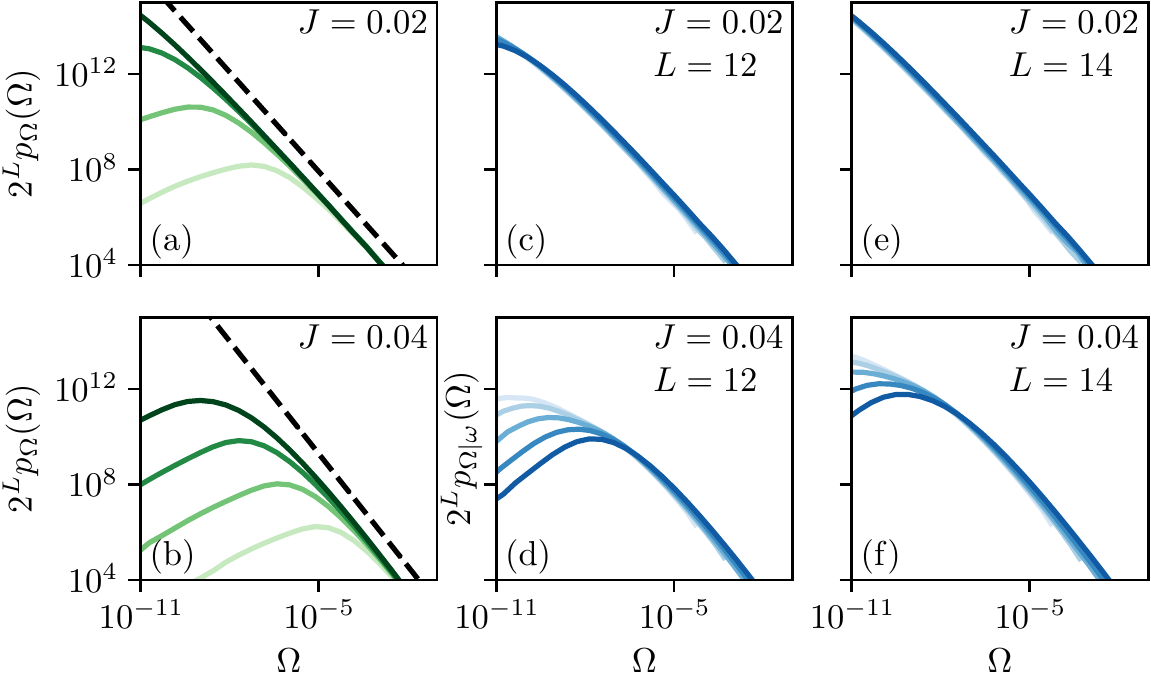}
	\caption{(a,b) Distribution $p_{\Omega}$ of $\Omega$ [Eq.~\ref{eq:Omega}] for $L=8,10,12,14$ (increasing from light to dark) with (a) $J=0.02$ and (b) $J=0.04$. Dashed lines indicate power-law fits to $p_{\Omega}$ for $L=14$, with exponents set by $\zeta/\zeta_c=0.39(1)$ for $J=0.02$ and $\zeta/\zeta_c=0.61(4)$ for $J=0.04$ [see Eq.~\ref{eq:pOmega}]. (c-f) Conditional distribution $p_{\Omega|\omega}(\Omega)$ of $\Omega < |\omega|$ within various windows of $|\omega|$. Windows are $|\omega| < 10^{-4}$ and $10^{-(n+1)} < |\omega| < 10^{-n}$ for $n=0,1,2,3$, with $|\omega|$ increasing from light to dark. 
	}
\label{fig:Omega}
\end{figure}

The correlations in Eq.~\eqref{eq:Omega} do not provide a prediction for $p_{\Omega}(\Omega)$, but they do provide a prescription for calculating it. Through exact diagonalisation (ED) it is straightforward to obtain the eigenstates and eigenvalues of the evolution operator, compute $z_{j,nm}$ and $\omega_{nm}$, and subsequently calculate the distribution of their product. We perform this analysis for a Floquet model for the MBL phase~\cite{lazarides2015fate,ponte2015manybody,zhang2016floquet} that was used previously in Refs.~\cite{garratt2021manybody,garratt2021local}. This model appears to be MBL for $J \leq 0.07$ and we provide details in Appendix~\ref{sec:model}. In Figs.~\ref{fig:Omega}(a,b) we show for small $J$ that $p_{\Omega}(\Omega)$ decays as a power of $\Omega$, and that for fixed $\Omega$ the distribution scales as $p_{\Omega}(\Omega) \propto 2^{-L}$. The exponential dependence on $L$ is necessary for the system to be MBL at large $L$, as we discuss in Sec.~\ref{sec:polarisations}. In Sec.~\ref{sec:counting} we rationalise the observed behaviour of $p_{\Omega}(\Omega)$ using a heuristic resonance counting argument. 

Equation~\ref{eq:Omega} has predictive power because, as argued in Sec.~\ref{sec:parametric}, the distribution of $\Omega$ is approximately independent of $\omega$ in the regime $\Omega_0 < \Omega < |\omega|$. Note that the lower cutoff $\Omega_0$ must be exponentially small in $L$ because $p_{\Omega}$ is proportional to $2^{-L}$ and decays as a power of $\Omega$. In order to confirm the weak correlations between $\Omega$ and $\omega$, in Figs.~\ref{fig:Omega}(c-f) we compute the ensemble-averaged conditional distributions $p_{\Omega|\omega}(\Omega)$ for various $\omega$. The collapse of $p_{\Omega|\omega}(\Omega)$ over several decades in both $\Omega$ and $\omega$ provides strong evidence that, when exploring the ensemble of MBL systems at a given $J$, variations of $z_{j,nm}\omega_{nm}$ and $\omega_{nm}$ are only weakly correlated. Since $p_{\Omega|\omega}(\Omega)$ is normalised, $\int^{\omega}_0 d\Omega p_{\Omega|\omega}(\Omega)=1$, small $\omega$ of course implies more weight at small $\Omega$, and this effect is clear in Figs.~\ref{fig:Omega}(c-f).

Comparing Eqs.~\ref{eq:Romegastar} and \ref{eq:Omega} it is clear that $\Omega_{j,nm}$ has a rough interpretation as the minimum splitting between a pair of levels passing through an avoided crossing. The statistics of these splittings have been studied in chaotic systems for some time \cite{wilkinson1989statistics,zakrzewski1991distributions,zakrzewski1993parametric2}, although an important contrast is that in our case the pairs of levels passing through avoided crossings are not necessarily nearest neighbours; it is clear from Fig.~\ref{fig:Omega} that there are resonant energy scales $\Omega$ that are much larger than the mean level spacing. We note that the energies associated with resonances on the scale of the system size $L$ (`end-to-end' resonances) were determined in Ref.~\cite{morningstar2021avalanches} by working with the many-body eigenstates themselves, as opposed to the matrix elements of local observables. There the approach is appropriate since for end-to-end resonances the appropriate two-dimensional subspaces are not locally embedded in many-body eigenstates, but are simply spanned by pairs of them. 

\section{Resonance counting}\label{sec:counting}
Here we provide a rationalisation of the behaviour in Fig.~\ref{fig:Omega}. Starting from a system of decoupled qubits, for small $J$ we suppose that it is possible to perform perturbation theory. Since this is a perturbation theory in a short-range interaction, we expect that each resonance can be associated with a length $r$, and that the corresponding energy scales $\sim e^{-r/\zeta}$, where the emergent quantity $\zeta$ is defined below (at small $J$, we expect ${e^{-1/\zeta} \propto J}$). If we choose to identify these energy scales with $\Omega$ defined in Eq.~\ref{eq:Omega}, we can estimate $p_{\Omega}(\Omega)$ via resonance counting. Note that for a given eigenstate and location in space, the number of possible resonances with length $r$ grows as $2^r$. Assuming that, for a given $r < L$, the various $\Omega$ are distributed between the respective lower and upper cutoffs $\Omega_0 \sim e^{-L/\zeta}$ and $e^{-r/\zeta}$, we find a power-law distribution
\begin{align}
	p_{\Omega}(\Omega) &= \frac{\zeta}{\zeta_c}\Omega^{\zeta/\zeta_c}_0 \Omega^{-(1+\zeta/\zeta_c)} \sim 2^{-L} \Omega^{-(1+\zeta/\zeta_c)},
\label{eq:pOmega}
\end{align}
which follows from weighting contributions from resonances with length $r$ by $2^r$. Here $\zeta_c \equiv [\ln 2]^{-1}$, ${\Omega_0 < \Omega < e^{-1/\zeta}}$, and $p_{\Omega}(\Omega) \to \delta(\Omega)$ as $J \to 0$. Note that the factor $2^{-L}$ here arose from the choice $\Omega_0 \sim e^{-L/\zeta}$. 

There is evident agreement between the counting argument leading to Eq.~\ref{eq:pOmega} and the exact numerics used for Fig.~\ref{fig:Omega}, so it is convenient to view Eq.~\ref{eq:pOmega} as the definition of $\zeta$. Both approaches lead to a power-law decay of $p_{\Omega}(\Omega)$ with $\Omega$, with exponent corresponding to $0 < \zeta/\zeta_c < 1$, and an exponential decay with $L$. Moreover, the values of $\zeta$ extracted from Figs.~\ref{fig:Omega}(a,b) [see caption] are consistent with the results of Ref.~\cite{garratt2021local}. From these estimates for $\zeta$ it can be verified that the power laws in Figs.~\ref{fig:Omega}(a,b) persist only down to $\Omega \simeq \Omega_0$, where the anticipated finite-size effects set in. 

\section{Local polarisations}\label{sec:polarisations}
Having established the behaviour in Eqs.~\ref{eq:Omega} and \ref{eq:pOmega}, we now discuss the statistical properties of the matrix elements $z_{j,nm}$ without any energy resolution. First, we calculate the distribution of $Z_{j,nm} = |z_{j,nm}|^2$, which will allow us to check whether our initial assumption, that $Z_{j,nm}$ is typically much smaller than unity, is consistent with the results in the previous sections. Following this we investigate the distribution of polarisations $|z_{j,nn}|^2 = 1-D_{j,n}$, where we have defined the depolarisation $D_{j,n}$ of $\ket{n}$ at site $j$. Because $\sigma_j$ squares to the identity, we have
\begin{align}
	D_{j,n} \equiv 1-Z_{j,nn} = \sum_{m \neq n} Z_{j,nm}.
\label{eq:sumrule}
\end{align}
Since a resonance between $\ket{n}$ and $\ket{m}$ at site $j$ is characterised by a matrix element $z_{j,nm}$ of order unity, the number of such contributions to the sum in Eq.~\eqref{eq:sumrule} is at most of order unity. For the MBL phase with $J \neq 0$ to resemble the $J=0$ limit we expect that typically $D_{j,n} \ll 1$, which is a far more stringent condition than $Z_{j,nm} \ll 1$.  

\begin{figure}
	\includegraphics[width=\linewidth]{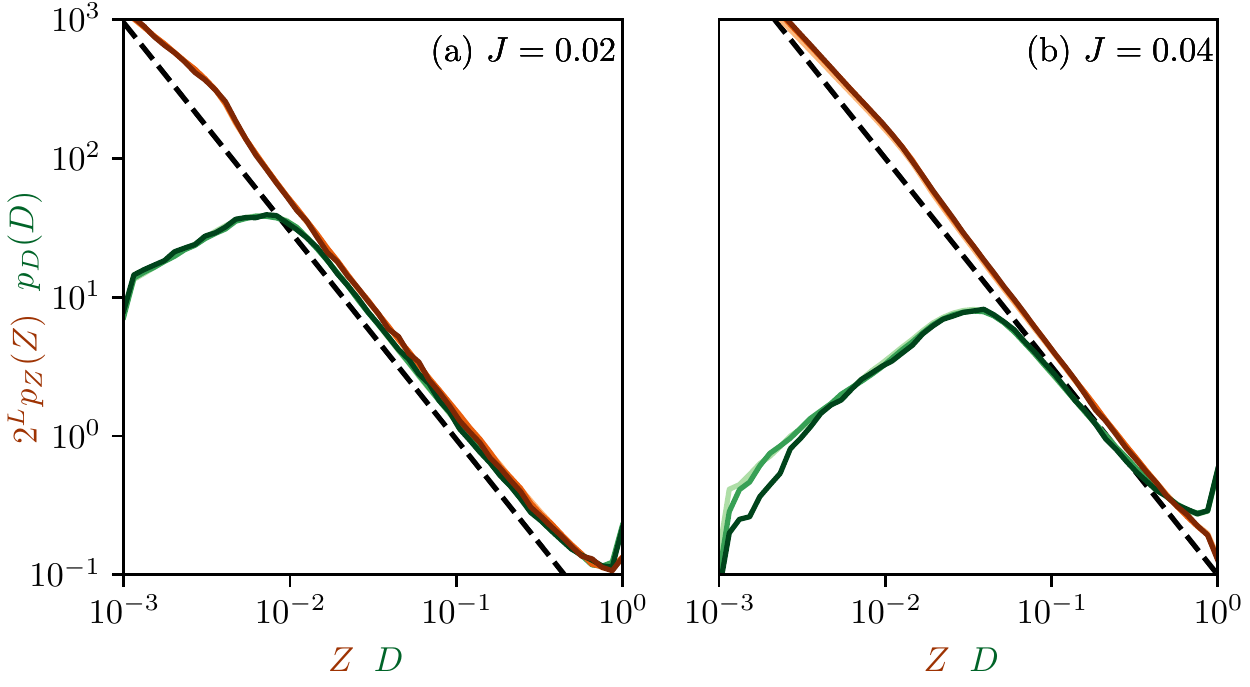}
	\caption{Distributions of modulus-squared off-diagonal matrix elements $Z_{j,nm} = |z_{j,nm}|^2$ (oranges) and depolarisations $D_{j,n}$ (greens) for (a) $J=0.02$, (b) $0.04$ and $L=8,10,12$ (light to dark). Dashed lines indicate the power-law decay $Z^{-3/2}$ expected from Eq.~\ref{eq:zdist}.}
	\label{fig:dz2}
\end{figure}

To understand the structure of Eq.~\ref{eq:sumrule} we compute the distributions $p_{Z}(Z_{j,nm})$ of $Z_{j,nm}$ and $p_D(D_{j,n})$ of $D_{j,n}$. The former is
\begin{align}
	p_{Z}(Z) = \Big\langle \delta\big( Z - \Omega^2/\omega^2 \big)\Big\rangle,
\end{align}
where the angular brackets denote an average over $\omega$ and $\Omega$. There is level repulsion on scales $|\omega| \sim \Omega$, but for $|\omega| \gg \Omega$ we can neglect correlations between $\omega$ and $\Omega$. Treating $\omega$ as uniformly distributed in this regime, we find from the Jacobian $\partial \omega /\partial Z \sim \Omega Z^{-3/2}$ [see Eq.~\ref{eq:Omega}] that for $\zeta<\zeta_c$ \cite{garratt2021local}
\begin{equation}
	p_{Z}(Z) \sim 2^{-L} g(\zeta) Z^{-3/2},\quad \Omega_0 \ll Z \ll 1, \label{eq:zdist}
\end{equation}
up to a prefactor of order unity. The function ${g(\zeta) = [\zeta_c-\zeta]^{-1} e^{[\zeta_c^{-1}-\zeta^{-1}]}}$ increases with $\zeta$ and diverges as $\zeta \to \zeta_c$. From Eq.~\eqref{eq:zdist} it is clear that $Z$ is typically much smaller than unity, as required for the approximation of decoupled resonances in Sec.~\ref{sec:parametric} to be appropriate. For example, the mean $\braket{Z} \sim g(\zeta) 2^{-L}$. As an aside we note that the distribution of the matrix elements of $\sigma_j$ has previously been characterised using an inverse participation ratio (IPR), which with exponent $q$ is defined as $\sum_{m} Z_{j,nm}^q$ \cite{monthus2016many,serbyn2017thouless}. From the distribution Eq.~\ref{eq:zdist} we see that the average IPRs are $L$-independent for $q > 1/2$.

From the above we find the average depolarisation $\braket{D} \sim g(\zeta)$. Crucially, because $p_{\Omega}(\Omega) \propto 2^{-L}$ and $\zeta<\zeta_c$, $\braket{D}$ is finite in the thermodynamic limit. This is of course essential for the existence of a MBL phase. Neglecting correlations between the different terms $Z_{j,nm}$ in Eq.~\ref{eq:sumrule} we can additionally calculate the distribution of depolarisations $p_D(D)$. When $D_{j,n}$ is dominated by just one term in the sum we have
\begin{align}
	p_D(D) \simeq 2^L p_{Z}(D) \sim g(\zeta) D^{-3/2},
\label{eq:Ddist}
\end{align}
In Fig.~\ref{fig:dz2} we compute $p_D(D)$ and $p_{Z}(Z)$ numerically, and find excellent support for Eqs.~\ref{eq:zdist} and \ref{eq:Ddist} for sufficiently large $Z$ and $D$. The result for $p_D(D)$ in Fig.~\ref{eq:Ddist} shows that the smallest polarisations are due to resonances, an $O(1)$ number of which dominate the sum in Eq.~\ref{eq:sumrule}. 

Interestingly, there is a `knee' in $p_Z(Z)$ in Fig.~\ref{fig:dz2}, below which $p_Z(Z)$ is larger than would be expected from Eq.~\ref{eq:zdist}. The origin of the knee can be understood by noting that, in calculating $p_Z(Z)$, there were two cases we should have distinguished. For half of the level pairs the maximum $Z$ on resonance is of order unity, and for the other half the maximum $Z$ on resonance should vanish as $J \to 0$, so for small $J$ we expect in the second case that the maximum $Z \sim J$. Another way to describe these two cases is to say that whereas in the first level repulsion restricts us to $|\omega| \gtrsim \Omega$, in the second we are restricted to $|\omega| \gtrsim \Omega/J$. Taking this into account in the calculation leading to Eq.~\ref{eq:zdist} gives rise to the observed enhancement of $p_Z(Z)$ for $Z \lesssim J$ relative to $J \lesssim Z \leq 1$. 

\section{Response to perturbations}\label{sec:perturbations}
Alongside the physical quantities discussed in the previous sections, another way to investigate resonances is by asking how many-body eigenstates respond to local perturbations. Here we focus on the fidelity susceptibility $\chi$. This quantity was introduced as a probe of singular behaviour at quantum quantum transitions \cite{sachdev2011quantum,zanardi2006ground,you2007fidelity}, and has more recently been investigated in the context of instabilities of the MBL phase and the onset of quantum chaos \cite{sierant2019fidelity,maksymov2019energy,sels2020dynamical,crowley2021constructive,pandey2020adiabatic,leblond2020universality,crowley2021partial}. Following this we revisit the level curvatures $\kappa$ discussed in Refs.~\cite{filippone2016drude,maksymov2019energy,garratt2021local}. In each case we will see that the statistical properties follow straightforwardly from the correlations in Eq.~\ref{eq:Omega}.

The fidelity susceptibility is here defined as ${\chi_{j,n} \equiv 4|\partial_{\lambda} \ket{n}|^2}$, where the additional factor of $4$ will be convenient below. From perturbation theory we have
\begin{align}
	\chi_{j,n} = \sum_{m \neq n} x_{j,nm}, \quad x_{j,nm} \equiv \frac{|z_{j,nm}|^2}{\omega_{nm}^2}. \label{eq:chix}
\end{align}
The distribution $p_{\chi}(\chi_{j,n})$ of $\chi_{j,n}$ is extremely broad, and consequently it is often more instructive to investigate this directly instead of to compute its moments. We will first compute the distribution $p_x(x_{j,nm})$ of $x_{j,nm}$, which is given by
\begin{align}
	p_x(x) =\Big\langle \delta( x - \Omega^2 \omega^{-4}) \Big\rangle,
\end{align}
where the angular brackets denote an average over $\Omega$ and $\omega$, and we have used Eq.~\ref{eq:Omega}. This is evaluated as a double integral over $\omega$ and ${\Omega < |\omega|}$, which proceeds as follows. For resonances on energy scale $\Omega$, the contribution to $p_x(x)dx$ from frequencies within $d\omega$ of $\omega = \Omega^{1/2} x^{-1/4}$ is $\sim \Omega^{1/2} x^{-5/4} dx$, where we have used the fact that the $\omega$ distribution is approximately uniform for $|\omega| > \Omega$. The restriction $\Omega < |\omega|$ corresponds to $\Omega < x^{-1/2}$, and integrating over resonances on all scales we have for $x \gg 1$,
\begin{align}
	p_x(x) \sim 2^{-L} x^{-5/4} \int_{\Omega_0}^{x^{-1/2}} d\Omega \, \Omega^{-(1/2 + \zeta/\zeta_c)},
\end{align}
from Eq.~\ref{eq:pOmega}. This expression highlights a change in behaviour at $\zeta/\zeta_c = 1/2$. Deep within the MBL phase with $\zeta/\zeta_c < 1/2$ the integral over $\Omega$ is dominated by resonances with $\Omega \sim x^{-1/2}$, and this leads to
\begin{align}
 	p_x(x) \sim 2^{-L} x^{-(3-\zeta/\zeta_c)/2}. \label{eq:px}
\end{align}
For $\zeta/\zeta_c < 1/2$ the distribution is controlled by pairs of levels close to resonance at $j$, in the sense that the off-diagonal matrix elements of $\sigma_j$ are large. For $\zeta/\zeta_c > 1/2$, on the other hand, the integral is dominated by contributions from resonances on the finest energy scales $\Omega \sim \Omega_0$ throughout the entire window $1 \ll x \ll \Omega_0^{-2}$. Note that this implies contributions from pairs of levels that may be very far from resonance at $j$, having $|\omega| \gg \Omega_0$. In this regime there is also a change in the scaling with $L$, and from Eqs.~\ref{eq:Omega} and \ref{eq:pOmega} we find $p_x(x) \sim e^{-L/(2\zeta)} x^{-5/4}$. Note, however, that Fig.~\ref{fig:Omega} as well as Ref.~\cite{morningstar2021avalanches} indicate that finite-size effects modify the distribution $p_{\Omega}(\Omega)$ relative to Eq.~\ref{eq:pOmega} for $\Omega \lesssim \Omega_0$, so in the following we restrict our analytic considerations to $\zeta/\zeta_c < 1/2$. 

Since $\chi$ is a sum of $\sim 2^L$ of the quantities $x$, it is straightforward to determine $p_{\chi}(\chi)$ in the regime where a single term dominates the sum. In that case $p_{\chi}(\chi) \simeq 2^L p_x(\chi)$. When $\zeta/\zeta_c < 1/2$ this occurs for $\chi$ larger than an $L$-independent threshold. We then find
\begin{align}
	p_{\chi}(\chi) \sim \chi^{-(3-\zeta/\zeta_c)/2}. \label{eq:pchi}
\end{align}
The slow decay of $p_{\chi}(\chi)$ with increasing $\chi$ shows clearly that the mean $\int d\chi \, p_{\chi}(\chi) \chi$ is controlled by resonances with $\Omega \sim \Omega_0$ \cite{sierant2019fidelity,sels2020dynamical,crowley2021constructive}. However, such resonances do not affect dynamics on physical time scales, highlighting the necessity of focusing instead on the full distribution.

The result Eq.~\ref{eq:pchi} illustrates the role of the correlations between the matrix elements $z_{j,nm}$ and the level separations $\omega_{nm}$. It should be contrasted with the distributions $p_{\chi}(\chi) \sim \chi^{-(3+\beta)/2}$ obtained for systems with Poissonian spectra ($\beta=0$) and random matrices drawn from the orthogonal ($\beta=1$) and unitary ($\beta=2$) ensembles \cite{sierant2019fidelity}. This dependence on the level-repulsion exponent $\beta$ can be seen to follow from (i) treating $z_{j,nm}$ as independent of $\omega_{nm}$ and (ii) the distribution of level separations $p_{\omega}(\omega) \sim |\omega|^{\beta}$ for $\omega$ below the mean level spacing. The results of this work, in particular those in Sec.~\ref{sec:energy}, clearly demonstrate that the first of these steps is inappropriate in MBL systems. Indeed, from Eq.~\ref{eq:pchi} we expect a decay of $p_{\chi}(\chi)$ slower than $\chi^{-3/2}$, but comparing with the random-matrix result $p_{\chi}(\chi) \sim \chi^{-(3+\beta)/2}$ appears to indicate an `effective' $\beta<0$. Neglecting the correlations between $z_{j,nm}$ and $\omega_{nm}$ could then lead to the erroneous conclusion $p_{\omega}(\omega) \sim |\omega|^{-\zeta/\zeta_c}$ at small $\omega$, i.e. `level attraction'~\cite{sels2020dynamical}, the absence of which is a central assumption in the proof of MBL in one dimension \cite{imbrie2016diagonalization}. Here we have shown that a decay of $p_{\chi}(\chi)$ slower than $\chi^{-3/2}$ arises in the case where there is only level repulsion, which becomes stronger for larger $\zeta/\zeta_c$ \cite{garratt2021local}. At the heart of this calculation is the fact that $z_{j,nm}$ and $\omega_{nm}$ are strongly correlated, and that these correlations are characterised the power-law distribution in Eq.~\ref{eq:pOmega}.

\begin{figure}
	\includegraphics[width=\linewidth]{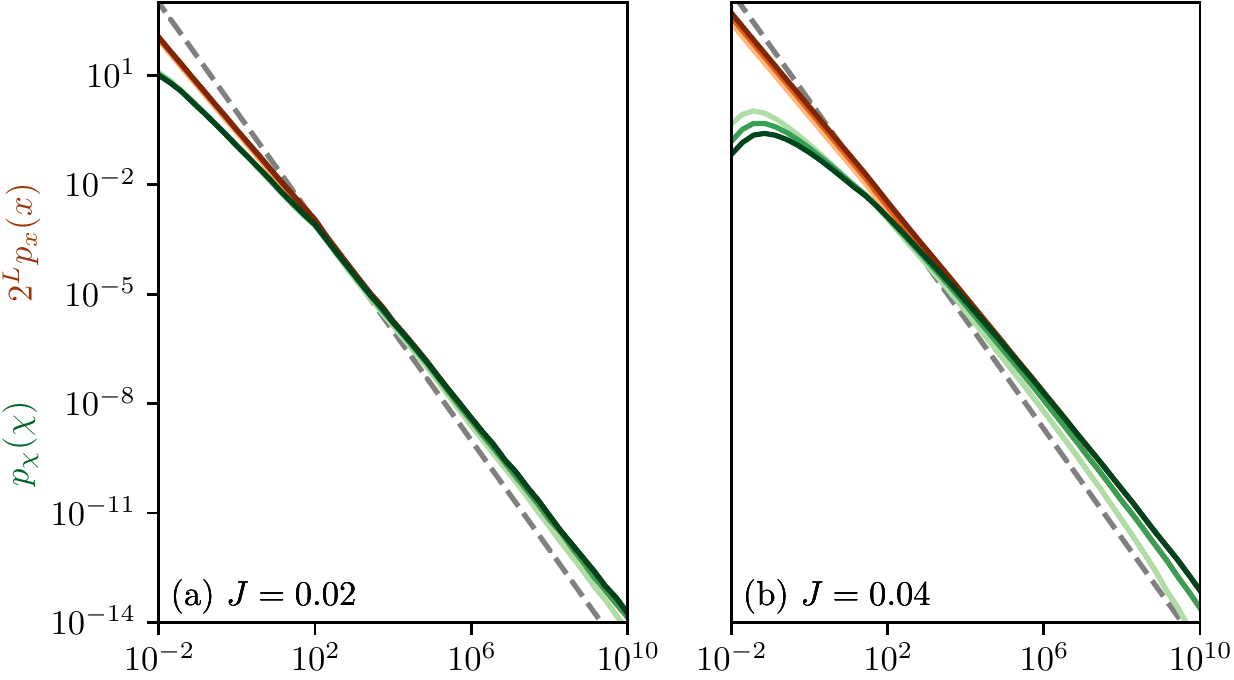}
	\caption{Distributions $2^L p_x(x)$ (oranges) and $p_{\chi}(\chi)$ (greens) for (a) $J=0.02$, (b) $0.04$ and $L=8,10,12$ (light to dark). Dashed lines indicate the power-law decay $\chi^{-3/2}$ expected for systems with Poissonian spectra.}
	\label{fig:chi}
\end{figure}

In Fig.~\ref{fig:chi} we compute $p_x(x)$ and $p_{\chi}(\chi)$ numerically. Since we perform these calculations for a Floquet model, the denominator $\omega^2_{nm}$ in Eq.~\ref{eq:chix} is replaced by $(2 \sin[\omega_{nm}/2])^2$, which follows from perturbation theory for Floquet operators as opposed to Hamiltonian ones. For large $\chi$ we find excellent agreement between $p_{\chi}(\chi)$ and $2^L p_x(\chi)$, showing that the sum $\chi_{j,n} = \sum_{m \neq n} x_{j,nm}$ is dominated by the largest term. The distributions determined numerically follow power laws over many decades in $\chi$, and the observed decay is marginally slower than $\chi^{-3/2}$, as expected from Eq.~\ref{eq:pchi}. 

Instead of discussing the overlaps of perturbed and unperturbed many-body eigenstates, we can also ask about changes in the expectation values of local observables. In particular, we can ask how $z_{j,nn}$ varies with the field $h_j$. It is an exercise in second order perturbation theory to show that the result is proportional to the level curvature $\kappa_{j,n}$. For convenience we define $\kappa_{j,n} = \partial_{\lambda} z_{j,nn}$ so that
\begin{align}
	\kappa_{j,n} = \sum_{m \neq n} k_{j,nm}, \quad k_{j,nm} \equiv \frac{|z_{j,nm}|^2}{\omega_{nm}},
\end{align}
and it is important to note that these quantities can have either sign. The calculation of $p_{\kappa}(\kappa)$ is similar to that of $p_{\chi}(\chi)$. For $\zeta/\zeta_c < 2/3$ the distribution $p_k(k)$ of $k$ is controlled by levels with $\omega \sim \Omega$, and so with $\Omega \sim |k|^{-1}$. For $\zeta/\zeta_c > 2/3$ it is controlled by $\Omega \sim \Omega_0$, and the $L$-dependence of the distribution is altered. For $\zeta/\zeta_c < 2/3$ we find
\begin{align}
	p_{\kappa}(\kappa) \sim |\kappa|^{-(2-\zeta/\zeta_c)}, \label{eq:pkappa}
\end{align}
previously derived in Ref.~\cite{garratt2021local}. The power law obtained there follows simply from the correlations in Eq.~\ref{eq:Omega} and the distribution of resonant energy scales in Fig.~\ref{fig:Omega} and Eq.~\ref{eq:pOmega}. 

Again we should contrast this distribution with the result for random matrices. For matrix elements of $\sigma_j$ that are independent of level separations, having $p_{\omega}(\omega) \sim |\omega|^{\beta}$ for small $\omega$, one instead finds $p_{\kappa}(\kappa) \sim |\kappa|^{-(2+\beta)}$ \cite{gaspard1990parametric,zakrzewski1993parametric1,vonoppen1994exact,fyodorov1995universality}. For each of $p_{\chi}(\chi)$ and $p_{\kappa}(\kappa)$, increasing the level repulsion exponent leads in the random-matrix setting to a faster power-law decay of the distribution, whereas in MBL systems increasing $\zeta/\zeta_c$ (and hence the degree of level repulsion) leads to a slower power-law decay. For large $\zeta/\zeta_c$, however, there is a distinct change in the distributions $p_{\chi}(\chi)$ and $p_{\kappa}(\kappa)$, which merits further investigation. 

\section{Summary}\label{sec:summary}
To summarise, we have formulated a theory for resonances in the MBL phase in terms of local observables. This allows for approximate analytic calculations that can be compared directly with numerics based on exact diagonalisation. A key ingredient is the relation between matrix elements and the many-body level spectrum, which we have argued can be understood based on parametric dynamics. This relation highlights the fact that signatures of resonance on energy scale $\Omega$ are evident even for level pairs separated in the spectrum by $|\omega| \gg \Omega$. Using our theory, we have shown how the resonant energy scales are encoded in the statistics of matrix elements, and this has allowed us to demonstrate numerically that they have a power-law distribution. This distribution should be contrasted with the one calculated in Ref.~\cite{morningstar2021avalanches} for pairs of states that are nearby on the scale of the mean level spacing. Our focus, as in Ref.~\cite{garratt2021local}, has been on finite energy scales, which corresponds to dynamics on finite time scales. 

We have additionally determined the form of the tail in the distribution of polarisations, which have been studied extensively in MBL spin chains since Ref.~\cite{pal2010many}. A complementary perspective comes from asking not about the statistical properties of individual systems, but instead asking how systems respond to local perturbation \cite{sierant2019fidelity,sels2020dynamical,crowley2021constructive}. In order to investigate this we have determined, both analytically and numerically, the distribution of eigenstate fidelity susceptibilities. Strikingly, this has uncovered a regime in which essentially the entire distribution is controlled by resonances on the finest energy scales, which are generally expected to correspond to extensive length scales. Our results also highlight an important difference relative to the behaviour of ergodic systems and random matrices, namely that increasing the degree of level repulsion in an MBL system leads to a slower power-law decay of this distribution rather than a faster one. The origin of this effect is in the extremely broad distributions of matrix elements of local observables. A similar contrast with random matrices is present when considering how eigenstate expectation values of local observables respond to perturbations. This is quantified in part by the distribution of level curvatures, calculated in MBL systems in Refs.~\cite{filippone2016drude,maksymov2019energy,garratt2021local}. In this work we have shown that this distribution, as well as the distribution of fidelity susceptibilities, follows from the correlations between matrix elements and the level spectrum in Eq.~\ref{eq:Omega}, and the power-law distribution of resonant energy scales in Eq.~\ref{eq:pOmega} and Fig.~\ref{fig:Omega}.

A contrast between our study and previous ones is that we have not described resonances in terms of LIOM \cite{gopalakrishnan2015low,crowley2021constructive,morningstar2021avalanches,garratt2021local}. Instead, we have worked only with quantities that can be calculated using standard numerical techniques. Our theory can therefore serve as a starting point for the numerical investigation of resonances closer to the transition to ergodic behaviour. Looking further afield, one might hope to extend the theory to understand the percolation of resonances, which is expected to drive the transition~\cite{deroeck2017many,deroeck2017stability,morningstar2021avalanches}.

\begin{acknowledgements}
We thank E.~Altman, V.~Bulchandani, D. A.~Huse and D.~E.~Logan for helpful discussions, J.~T.~Chalker for guidance and collaboration on related work~\cite{garratt2021local}, and A. Chandran, M. Fava and F. Machado for useful comments on the manuscript. This work was in part supported by the Gordon and Betty Moore Foundation (SJG), a  ICTS-Simons Early Career Faculty Fellowship (SR) and EPSRC Grant No. EP/S020527/1 (SR).
\end{acknowledgements}

\appendix

\section{Floquet Model}\label{sec:model}

Here we describe the Floquet model for the MBL phase used for numerical calculations. In our models all points in the spectrum of the Floquet operator $W$ are statistically equivalent, and there are no conserved densities. For integer time $t$ the unitary evolution operator is $W^t$, and $W\ket{n} = e^{i \theta_n}\ket{n}$ where the quasienergies $\theta_n \in [-\pi,\pi)$. The Floquet operators that we use have the structure of brickwork quantum circuits, specifically $W = [\bigotimes_{j\text{ odd}} w_{j,j+1}][\bigotimes_{j\text{ even}} w_{j,j+1}]$ where $w_{j,j+1} = \exp\big[i\pi J\Sigma_{j,j+1}\big]\big[u_{j}\otimes v_{j+1}\big]$. 
Here $\Sigma_{j,j+1}$ is the swap operator, or equivalently a Heisenberg coupling, acting on qubits $j$ and $j+1$, while $J$ is the coupling strength. 
The independent Haar-random $2 \times 2$ unitary matrices $u_j$ and $v_j$ describe the random fields, and due to these fields our model does not have time-reversal symmetry. Up to an overall phase, we can write $v_j u_j$ ($u_j v_j$) for $j$ even (odd) as $e^{i \vec{h}_j \cdot \vec{\sigma}_j/2} \equiv e^{i h_j \sigma_j/2}$, where $\vec{\sigma}_j$ is a vector of Pauli matrices and $h_j \equiv |\vec{h}_j|$ so that $\sigma_j$ has eigenvalues $\pm 1$. We restrict ourselves to behaviour deep within the MBL phase, and so with $J$ well below $0.07$~\cite{garratt2021manybody}. Note that with periodic boundary conditions the structure of the evolution operator necessitates $L$ even. 

\section{Solution of parametric equations}\label{sec:solution}

Here we outline the solution of the system of equations in Eq.~\ref{eq:dz}. For brevity we drop indices, e.g. $z = z_{j,nm}$, and shift the phase of $z$ so that it is real. Then  
\begin{align}
	\partial_{\lambda} \omega = y, \quad 
	\partial_{\lambda} y = \omega^{-1} z^2, \quad
	\partial_{\lambda} z = -\omega^{-1} y z\,, 
	\label{eq:diffeq}
\end{align}
which describe an avoided crossing in a two-dimensional resonant subspace. In contrast with Ref.~\cite{garratt2021local} it has not been necessary to introduce operators that describe the dynamics of subsystems, although Eqs.~\ref{eq:diffeq} can also be derived using that approach. As discussed in the main text, solutions to Eq.~\ref{eq:diffeq} are parametrised by three constants of integration. Two of these are $\lambda^*$ and $\omega^*$, which can be respectively identified with the fictitious time and splitting at the resonance. The third constant of integration is $R^2 \equiv y^2+z^2$. Note that for $J=0$ we necessarily have $z=0$, while $y^2 = 0$ or $1$, corresponding to level pairs that have the same or opposite polarisations at the site of the perturbation. For small $J \neq 0$ we therefore expect level pairs to have either $R^2$ of order $J$, or $R^2 = 1-O(J)$. 

To solve Eqs.~\ref{eq:diffeq} it is convenient to write $y(\lambda)  = -R\cos\varphi(\lambda)$ and $z(\lambda) = R\sin \varphi(\lambda)$. This leads to
\begin{align}
\begin{split}
	\tan \varphi(\lambda) &= \frac{\omega^*}{R(\lambda^*-\lambda)},  \\ \omega^2(\lambda) &= R^2 (\lambda-\lambda^*)^2+(\omega^*)^2, 
	\end{split}
	\label{eq:tanphi}
\end{align}
which gives matrix elements of $\sigma_j$ set by
\begin{align}
	y(\lambda) = \frac{R(\lambda^\ast-\lambda)}{\omega(\lambda)}, \quad z(\lambda) = \frac{R\omega^*}{\omega(\lambda)}.
\label{eq:pairwiseyz}
\end{align}
Setting $\lambda=0$ in the second of Eqs.~\ref{eq:pairwiseyz} we arrive at Eq.~\ref{eq:Omega}, and identify $\Omega = R \omega^*$. In the case $R^2=1-O(J)$ the energy scale $\Omega$ that controls the off-diagonal matrix elements of the local observable $\sigma_j$ is approximately equal to the minimum level separation $\omega^*$ that arises under variations in $\lambda$. 

\begin{figure}
	\includegraphics[width=\linewidth]{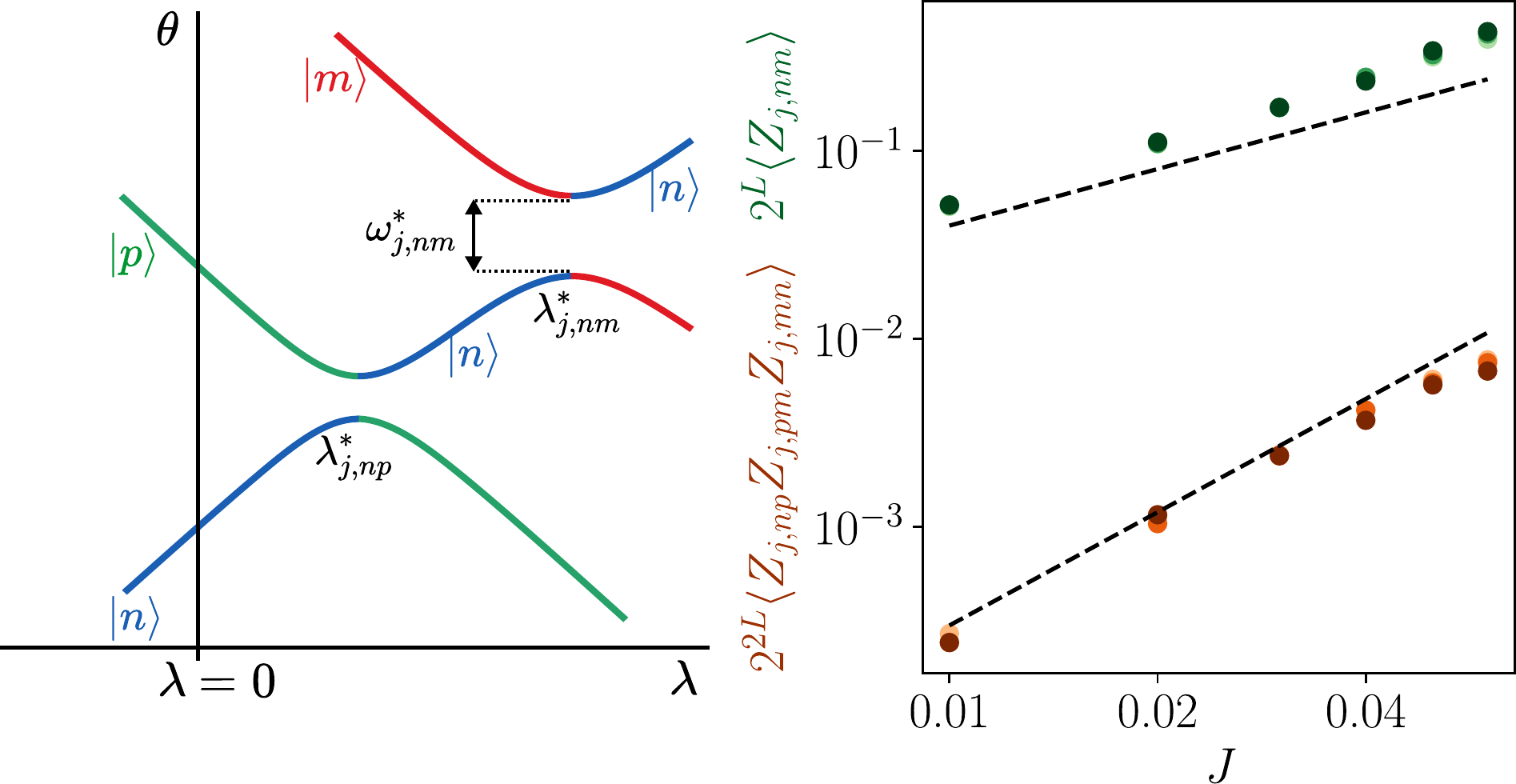}
\caption{Left: Diagram of parametric dynamics with three levels, with quasienergy $\theta$ vertical and fictitious time $\lambda$ horizontal. The exchange of labels at avoided crossings is described after Eq.~\ref{eq:pairwiseyz}. Right: Comparison of (greens) two-level $\braket{Z_{j,nm}}$ and (oranges) three-level $\braket{Z_{j,np}Z_{j,pm}Z_{j,mn}}$ terms for $L=8,10,12$ (increasing from light to dark). Dashed lines indicate growth with $J$ and with $J^2$.}
\label{fig:threebody}
\end{figure}

\section{Variations of $\Omega$}\label{sec:threebody}
Moving beyond a two-dimensional resonant subspace, here we consider applying our theory to the full spectrum; aspects of this are discussed elsewhere \cite{garratt2021local}. First, it is necessary to adopt a convention for the relabelling of levels. The simplest choice is to exchange the labels of levels at the centres of avoided crossings, as illustrated in Fig.~\ref{fig:threebody}(a). In this way the polarisations $z_{j,nn}$ have the same signs on both sides of these crossings. Second, we must ask how and whether $\Omega_{j,nm} = |\omega_{j,nm} z_{j,nm}|$ varies with $\lambda$ once we account for coupling to levels $\ket{p} \neq \ket{n},\ket{m}$.

This coupling can be analysed within perturbation theory in $\lambda$. Variations of the energy scale $\Omega_{j,nm}$ are given by
\begin{align}
\begin{split}
	\partial_{\lambda} \Omega_{j,nm}^2 &= \sum_{p \neq n,m} f_{j,nmp}(z_{j,np} z_{j,pm} z_{j,mn} + \text{c.c.})\,,\\
	f_{j,nmp} &= \frac{1}{2}\omega^2_{j,nm}[ \omega_{j,mp}^{-1} + \omega_{j,np}^{-1} ]\,.
	\end{split}
\label{eq:dOmegadlambda}
\end{align}
In order for Eq.~\ref{eq:Omega} to be appropriate even for ${|\omega_{nm}| \gg \Omega_{j,nm}}$, the net variation in the resonant energy scale $\Omega_{j,nm}$ from fictitious time $\lambda^*_{j,nm}$ to the random realisation $\lambda=0$ of interest must be small. To understand how this can be the case, it is simplest to consider a level $\ket{p}$ with $R^2_{j,np}$ of order unity. This scenario is illustrated on the left in Fig.~\ref{fig:threebody}. Note that although the factor $\omega_{j,pn}^{-1}$ appearing in $f_{j,nmp}$ can be large, due to the above relabelling scheme it has opposite signs for $\lambda < \lambda^*_{j,np}$ and $\lambda > \lambda^*_{j,np}$. This means that a decrease in $\Omega_{j,nm}^2$ on one side of the resonance between $\ket{n}$ and $\ket{p}$ is (approximately) compensated by an increase on the other side. The net variation in $\Omega_{j,nm}$ is therefore suppressed.

An interesting effect beyond our two-level approximation appears in the quantity $z_{j,np} z_{j,pm} z_{j,mn}$ in Eq.~\ref{eq:dOmegadlambda}, and we now briefly discuss its modulus square, $Z_{j,np}Z_{j,pm}Z_{j,mn}$. In particular, these three matrix elements are not independent of one another. Suppose, as above, that $R^2_{j,mn}$ and $R^2_{j,np}$ are both of order unity while $R^2_{j,pm}$ is of order $J$. For such a set of levels we have $\braket{Z_{j,mn}},\braket{Z_{j,np}} \sim 2^{-L} J$ while $\braket{Z_{j,pm}} \sim 2^{-L}J^2$. If we neglect correlations between these matrix elements, the disorder-average $\braket{Z_{j,np}Z_{j,pm}Z_{j,mn}} \sim 2^{-3L} J^4$. However, if there is a resonance between $\ket{n}$ and each of $\ket{m}$ and $\ket{p}$, this causes $Z_{j,pm}$ to acquire a value of order unity. As a consequence we instead have $\braket{Z_{j,np}Z_{j,pm}Z_{j,mn}} \sim 2^{-2L} J^2$, and in Fig.~\ref{fig:threebody} we confirm this behaviour numerically. The $J^2$ scaling nevertheless reflects the rarity of three-level relative to two-level resonances.

%

\end{document}